# Selection Against Demographic Stochasticity in Age-Structured Populations

Max Shpak, Department of Biological Sciences, University of Texas at El Paso, El Paso TX 79968, USA,
Ph: (915)747-8903, fax: (915) 747-5808, mshpak@utep.edu

## Abstract

It has been shown that differences in fecundity variance can influence the probability of invasion of a genotype in a population, i.e. a genotype with lower variance in offspring number can be favored in finite populations even if it has a somewhat lower mean fitness than a competitor. In this paper, Gillespie's results are extended to population genetic systems with explicit age structure, where the demographic variance (variance in growth rate) calculated in the work of Engen and colleagues is used as a generalization of "variance in offspring number" to predict the interaction between deterministic and random forces driving change in allele frequency. By calculating the variance from the life history parameters, it is shown that selection against variance in the growth rate will favor a genotypes with lower stochasticity in age specific survival and fertility rates. A diffusion approximation for selection and drift in a population with two genotypes with different life history matrices (and therefore, different growth rates and demographic variances) is derived and shown to be consistent with individual based simulations. It is also argued that for finite populations, perturbation analyses of both the growth rate and demographic variances may be necessary to determine the sensitivity of "fitness" (broadly defined) to changes in the life history parameters.

**Introduction**

Theoretical population genetics deals primarily with furthering the understanding of how stochastic and deterministic factors interact to drive change in gene frequency p in a population. The deterministic force in question is usually natural selection (on survivorship, fecundity, etc) while the stochasticity can arise from either extrinsic or intrinsic forces. In the former case, it can be due to fluctuating environmental conditions (Haldane and Jayakar 1955, Lewontin and Cohen 1969, Tuljapurkar and Orzack 1980), or to the familiar process of finite sampling of gametes from a large gamete pool that leads to the standard scenarios of genetic drift (e.g. Wright 1931, Kimura 1964). In contrast, intrinsic variance in fitness arises from the fact that survival and reproductive output are themselves probabilistic events, where even in a constant environment, a given genotype will not produce a fixed number of offspring or survive until the same age.

Gillespie (1974, 1975) investigated the role of intrinsic variance in offspring number, showing that the contribution of this variance to the change in allele frequency arises as a consequence of sampling from a finite gamete pool, as opposed to sampling of gametes from an effectively infinite gamete pool in Fisher-Wright-Kimura models of genetic drift (furthermore, when the fertility distribution is Poisson, the variance contribution is of the order 1/N and effectively negligible). Gillespie's results demonstrated that in a finite population, the variance contributes to both the first and second moments of Δp=p(t+1)-p(t). This is in contrast to Wright-Kimura genetic drift, which does not influence E[Δp] but only the expectation $E[(\Delta p)^2]$.

Specifically, for two competing asexual haploid genotypes in a population of size N, with respective average and variance in offspring number $\mu_1$, $\mu_2$, $\sigma_1^2$, $\sigma_2^2$, the first and second moments for the change in frequency p of the first genotype are, to a first order approximation when the variance are small and the averages are near unity,

$$E[\Delta p] \approx p\,(1-p)\left(\Delta\mu - \frac{\Delta\sigma^2}{N}\right) \qquad (1\,a)$$

$$E[\Delta p^2] \approx \frac{p\,(1-p)}{N}\,((1-p)\,\sigma_1^2 + p\,\sigma_2^2) \qquad (1\,b)$$

(see also Proulx 2000 for a different derivation of these expressions).

It can be seen from (1.a) that in a finite population, directional selection will favor reduced offspring number (fitness) variance by tending to increase the frequency of genotypes with lower $\sigma^2$ values. In particular, when the mean fitness differences are relatively small, selection will tend to favor the genotype with lower variance. In fact, selection can actually favor a genotype with lower average fitness if it sufficiently reduces the variance, i.e. when $\Delta\mu < \frac{\Delta\sigma^2}{N}$, the allele with lower variance will have a higher than neutral probability of fixation. Intuitively, the fact that fixation probabilities depend on the variance as well as the first moment in fitness can be understood from the standpoint of "bet hedging" (Seger and Brockman 1982, Stearns 2000). Even if the mean fitness is high, a high variance value will lead to few (or no) offspring in some generations, thus leading to lineage extinction. This effect is more pronounced in a small population (since variance in a sample mean parameter scales inversely with sample size N), since for N large at least some individuals are likely to compensate with many offspring for those which have few.

Under the assumptions of weak selection and low variance, 1a and b can be used, respectively, to approximate the drift and diffusion coefficients in the Kolmogorov backward equations. These give estimates for fixation probabilities of a given genotype when the only source of stochasticity is intrinsic, i.e.

$$U(p) = \frac{\int_0^p ((1-x)\sigma_1^2 + x\sigma_2^2)^{2\left(\frac{N(\Delta\mu)}{\Delta\sigma^2} - 1\right)} dx}{\int_0^1 ((1-x)\sigma_1^2 + x\sigma_2^2)^{2\left(\frac{N(\Delta\mu)}{\Delta\sigma^2} - 1\right)} dx} \quad (2)$$

which is independent of population size when the mean fitness difference is 0. The predictions of the diffusion approximation for selection on fertility variance are consistent with simulation results (Gillespie 1974, Shpak 2005). Strictly speaking, the above expression is an approximation, based on a continuous time diffusion representation of (1). As such, it requires that $\mu \approx 1+r$ with r small for both genotypes, since a representation in continuous rather than discrete time is in terms of log scale transformations of the parameters (see below), with r=Log[$\mu$] in place of $\mu$ and $\sigma^2/\mu^2$ as the variance term to give appropriate scaling of the variance terms.

Gillespie's presentation of selection on variance in offspring number does not deal explicitly with age structure (i.e. in the derivations, there is effectively a single age class with variance in fertility) or any other life history variable. In his papers, it was assumed that the number of offspring contributed by a genotype was a random variable described as $\mu+\delta$, where $\delta$ was a random variable with variance $\sigma^2$. The variance would arise as a consequence of the genotype undertaking different life history "strategies," although the variances was never explicitly derived in these terms. Demetrius and Gundlach (2000) rederived the results of Gillespie for a model where fecundity is a random variable associated with the occurrence of a probability distribution of strategies and the variance is explicitly defined in terms of this distribution.

Age structure was implicit in an example used to motivate the question, however. In Gillespie (1974), divergent variances in offspring number were proposed to arise from semelparous versus iteroparous life histories. Specifically, if one genotype produces a single clutch of k offspring that survive or die (collectively) with probability $\pi$, while another produces k individual offspring at different points, each of which survives individually with probability $\pi$. While both strategies have the same expected number of offspring, $E[x]=k\pi$, the variance in surviving offspring number in the former, semelparous case is $Var[x]= k^2\pi(1-\pi)$ versus the smaller $k\pi(1-\pi)$ in the iteroparous scenario.

Substituting these binomial variance values into equations (1) and (2) gives an estimate of fixation probabilities that is in close agreement with the frequencies of fixation or loss in individual-based simulations of these life histories under a variety of conditions (Shpak 2005), at least for relatively small selection coefficients. However, it should be apparent that substituting these values is only an approximation, in that it reduces an evolutionary model with age structure (for the iteroparous individual) to a unidimensional one without age classes. Consequently, selective dynamics driven by differences in generation time or in reproductive value contributions from each age class (under potentially different distributions) are ignored, with potentially misleading results. For a complete understanding of how selection for variance in offspring number operates, the variance terms must be calculated explicitly from models with age structure. In the simple example of semelparity and iteroparity outlined above as in more general models, the variance can be derived from Leslie matrix representations.

Below, it will be shown that the variance calculations from life history matrices can be used directly as terms in (1), (2). Much of the presentation and notations follows the work of Steinar Engen and colleagues, with the intention of generalizing their results to competition between two different life histories, in which there may be trade-offs between maximizing mean and minimizing the variance in growth rate across strategies.

## Demographic Variance in Age Structured Populations

For a vector $\mathbf{n}=\{n_1...n_d\}$ describing the number of individual $x_i$ in the ith age class, the number of individuals in the next time step can be predicted from the fecundities and survival frequencies of each class. Leslie (1945) introduced a matrix representation such that $\mathbf{n}(t+1)=\mathbf{L}\mathbf{n}(t)$,

$$\mathbf{L} = \begin{pmatrix} F_1 & F_2 & \cdots & \cdots & F_d \\ b_1 & 0 & 0 & 0 & 0 \\ 0 & b_2 & 0 & 0 & 0 \\ 0 & 0 & \ddots & 0 & 0 \\ 0 & 0 & 0 & b_{d-1} & 0 \end{pmatrix}$$

where $F_i$ is the fertility of the ith age class and $b_i$ the fraction of individuals in that class that survive to the next time step. Treated as a deterministic dynamical system where the parameters are rates, the population dynamics and evolution of age structured systems is well studied and understood (e.g. Charlesworth 1994, Caswell 2001). Much of the theory is based on an understanding of the eigenstructure of $\mathbf{L}$, where the leading eigenvalue $\lambda$ gives the growth rate when the age class distribution is close to the frequencies given by the corresponding eigenvector.

However, a strictly deterministic interpretation of the Leslie matrix obscures the fact that the parameters b and F are actually random variables (so that deterministic models based on Leslie matrices actually use the mean values $\bar{b}_i$ and $\bar{F}_i$ as their coefficients for each age class parameter). In natural populations, it is not a fixed fraction $b_i$ of individuals that survive to the i+1 age class, because survival of any individual from year to year itself probabilistic. Rather, $b_i$ is a Bernoulli probability with mean $b_i$ and variance $b_i(1-b_i)$. This in turn implies that the number of survivors in the next age class out of a pool of $n_i$ will have a binomial distribution with mean $n_i b_i$ and variance $n_i b_i(1-b_i)$, while the variance in the fraction of survivors is $b_i(1-b_i)/n_i$. The fertility values in any given age class $F_i$ are also (often) themselves random variables, since the reproductive output of a given age class (even holding genotype constant) is seldom fixed. If the census point of the population is taken prior to reproduction, the fertility of an age class is the product of that age class's fecundity $m_i$ multiplied by the probability of surviving through to the end of the first age class, i.e. $F_i = b_o m_i$. Thus, even when $m_i$ is assumed constant, there is again a contribution from the Bernoulli variance $b_o(1-b_o)$ of survival probabilities (Caswell 2001).

If $m_i$ is itself a random variable (which can follow virtually any probability distribution, characterized by mean $m_i$ and variance $\sigma_i^2$), one has true "variance in offspring number," and in the limiting case of a single age class, one has the variance quantity seen in Gillespie (1975), where $\sigma_i^2$ is the variance term in (1) and (2). Note that in an age structured population, it is not accurate to describe the variance that arises from differential survivorship across age classes as "variance in offspring number" and to use it as anything but an approximation in Gillespie's equations. In order to correctly describe selection on variance in age structured populations in terms of a single variance parameter, one needs to calculate not the variance in offspring number (fecundity) but the rather variance in the population growth rate.

This quantity, referred to as demographic variance, has been derived and discussed in detail in Caswell (2001, Ch. 15) and in the works of Steinar Engen and coworkers (e.g. Engen et al 2005a,b, Lande et al 2003). Following the presentation in Lande et al (2003), if $\lambda$ is the growth rate, one can write $\lambda = \mu + \delta$, where $\mu = E[\lambda]$ is the expected growth rate of a single genotype and $\delta$ is a random variable with mean 0 and a variance of $\sigma_d^2$. In a population of N individuals, variation in growth rate is due to the variance of the sample mean, i.e.

$$\text{var}[\lambda] = \sigma^2 = \text{Var}[\mu] + \text{Var}\left[\frac{1}{N}\sum_{i=1}^{N}\delta_i\right] = \frac{\sigma_d^2}{N} \qquad (3)$$

(since $\mu$ is a constant).

With population size on a log scale X=log[N], the growth rate r=log[$\lambda$], which from a Taylor expansion of $\lambda$ about the mean is

$$r = \text{Log}[\mu + \delta] = \text{Log}\left[\mu\left(1 + \frac{\delta}{\mu}\right)\right] \approx \text{Log}[\mu] - \frac{\sigma^2}{2\mu^2} = \bar{r} - \frac{\sigma_d^2}{2\mu^2 N} \qquad (4)$$

(here $\sigma^2$ is the variance in sample mean growth rate, which is simply the demographic variance $\sigma_d^2/N$). For a single genotype, the growth rate in a finite population is estimated as the difference between mean growth rate and the (suitably scaled) demographic variance divided by population size. It will be seen below that in competition between different genotypes this will be the relevant measure of fitness.

The demographic variance, in turn, can be calculated directly from the Leslie matrix parameters when one assumes that $F_i$ and $b_i$ are themselves random variables. To calculate the variance of $\lambda$, we note that for any function $f(x_1, x_2...x_k)$ of random variables $x_i$, the variance of f(**x**) up to second order approximation is given by

$$\text{Var}[f(x_1, x_2 \ldots x_k)] \approx \sum_{i,j} \frac{\partial f}{\partial x_i} \frac{\partial f}{\partial x_j} \text{Cov}[x_i, x_j]$$

which in the absence of covariances between parameters,

$$\text{Var}[f(x_1, x_2 \ldots x_k)] \approx \sum_i \left(\frac{\partial f}{\partial x_i}\right)^2 \text{Var}[x_i]$$

Since the growth rate λ (and therefore, the Malthusian parameter r) is a function of the Leslie matrix parameters, the variance in λ can be estimated as (Caswell 2001, Lande et al 2003, Engen et al 2005a,b),

$$\sigma_d^2 = \sum_i \left(\frac{\partial \lambda}{\partial x_i}\right)^2 \frac{\text{Var}[x_i]}{n} \tag{5}$$

The relevant parameters are $b_i = \bar{b}_i + \epsilon_1$, $F_i = \bar{F}_i + \epsilon_2$. If survival of each individual in an age class is assumed to be a Bernoulli random variable, $\text{var}(b_i) = b_i(1-b_i)$, while the variance in $F_i$ is the combined contribution of variance in offspring number $m_i$ and the survival probability from birth to the first age class $p_0$. For a population in age class equilibrium $\nu$, the variance of the sample mean contribution of each age class parameter, $n_i$ is the product of census population size and the equilibrium frequency of the ith age class, $N\nu_i$.

Furthermore, the partial derivatives of the growth rate with respect to the life history parameters have been calculated (Demetrius 1969, elaborated in Caswell 2001), as

$$\frac{\partial \lambda}{\partial b_i} = \frac{\lambda^{-i} l_i V_{i+1}}{\tau}, \quad \frac{\partial \lambda}{\partial F_i} = \frac{\lambda^{-(i-1)} l_i}{\tau}$$

where $l_i$ is the probability of survival to age class i, $l_i=\prod_{k=0}^{i-1} b_k$, while $\tau$ and V refer, respectively, to the expected generation time of a life history specified by Leslie matrix **L**, and the reproductive value (sensu Fisher 1958) of an individual in the ith age class (i.e. the expected number of progeny of an individual in age i from the present time until its death). These quantities are

$$\tau = \sum_{k=1}^{d} k \lambda^{-k} l_k F_k$$

$$V_i = \frac{\lambda}{l_i} \sum_{k=i+1}^{d} \lambda^{-k} l_k F_k$$

Putting the above expressions together, one has an estimate for demographic variance which can be calculated directly from the life history parameters (Engen et al 2005):

$$\sigma_\lambda^2 = \sum_i \left[ \left(\frac{\partial \lambda}{\partial F_i}\right)^2 \text{var}[F_i] + \left(\frac{\partial \lambda}{\partial b_i}\right)^2 \frac{b_i(1-b_i)}{V_i} \right]$$

$$= \sum_i \left[ \left(\frac{\lambda^{-i+1}}{\tau}\right)^2 \text{var}[F_i] + \left(\frac{\lambda^{-i} l_i V_{i+1}}{\tau}\right)^2 \frac{b_i(1-b_i)}{V_i} \right] \quad (6)$$

On the logarithmic scale, the demographic variance is $\sigma_d^2 = \sigma_\lambda^2/\mu^2$.

It was demonstrated by Engen et al (2007) that strictly speaking, the above growth and variance terms do not necessarily describe first and second moments in the population growth rate in the number of individuals from one generation from the next. Rather, the growth rate $\lambda$ determines the rate of increase in the population reproductive value, which is given by V=**V.n**=$\sum_i V_i n_i$ (where **V** is the vector of reproductive values and **n** a vector describing the number of individuals in each age class) and from the fact that the distribution of reproductive values is the left eigenvector of **L**, we have E[V(t+1)|V(t)]=$\lambda$V(t). Because of the stochasticity inherent in the survival and reproductive processes, the distribution of age classes from one time step to the next will deviate from the stable age frequency *v*. This deviation is especially evident when the generation time of the life history is significantly longer than a single time step (Engen et al, 2007).

As a result, the growth rate and its variance will only approximately describe the first and second moments in N(t) from one time step to the next. While the expected growth rate of N(t) is the same as for V(t), the variance of the former will actually be

$$\sigma_N^2 = \sum_i v_i \left( \text{var}[F_i] + b_i (1 - b_i) \right) \qquad (7)$$

i.e. the variance contribution of each age class weighted by its frequency near equilibrium. It can be seen that equation (7) corresponds closely with (6) only when $\tau \approx 1$ and when the reproductive value is approximately equal to the fertility of the age class in question. However, over multiple time steps (t>>$\tau$), the approximation N(t)$\approx\lambda^t$V(0)$\approx\lambda^t$N(0). Consequently, demographic variance of reproductive value can be used as a reliable estimate for the second moment of population growth rate given sufficient time.

In formal terms, the convergence of first and second moments in population growth rate to the dynamics of reproductive value is given by

$$r = \log[\lambda] = \lim_{t \to \infty} \frac{1}{t} \text{Log}\left[\frac{N(t)}{N(0)}\right]$$

$$\sigma_r^2 = \lim_{t \to \infty} \frac{1}{t} \text{Var}\left(\text{Log}\left[\frac{N(t)}{N(0)}\right]\right)$$

The approximation of mean and variance in reproductive value growth rate describing the first and second moments in population number is useful for a number of reasons. First, population census number is easier to estimate than population reproductive value. More importantly, describing the system in terms of N(t) is critical for deriving equations describing competition between two genotypes with different life history parameters. Engen et al (2005b) do so for the special case of equal average growth rate and equal demographic variances (i.e. no selection on either the growth or variance parameters), but the results readily generalize to different mean values to give equations in the same form as those presented by Gillespie, with mean growth rate in place of mean offspring number and demographic variance in place of variance in offspring number.

## Diffusion Approximations: Combining Deterministic and Stochastic Forces

For N individuals of a given allele (or genotype) in a population of asexual haploids, the change in population number n(t) with time can be represented with the stochastic differential equation in the following form,

$$d n(t) = M(n) \, dt + \sqrt{S(n)} \, dB_t \tag{8}$$

where $dB_t$ describes standard Brownian motion. The coefficients M(n) and V(n), respectively, are the first and second moments in the change in the population number of the first genotype,

$$M(n) = \lim_{\Delta t \to 0} \frac{1}{\Delta t} E[n(t + \Delta t) - n(t) \mid n(t) = n],$$

$$S(n) = \lim_{\Delta t \to 0} \frac{1}{\Delta t} E[(n(t + \Delta t) - n(t))^2 \mid n(t) = n]$$

Here M(n)=rn and S(n)=$\sigma_r^2 n$, which follow from an appropriate change of variable from the log scale (see Engen et al 2005, pg 951). From a Taylor series expansion of Log[$\lambda$], $\sigma_r^2 = \frac{\sigma_\lambda^2}{\lambda^2}$. In order for the above differential approximation to hold, it must be the case that $\lambda \approx 1+r$, otherwise terms of higher than second order in the expansion of r=log[$\lambda$] cannot be ignored.

Equation (8) can be equivalently expressed as a Kolmogorov forward equation, where the probability density function of n as $\phi$(n,t)

$$\frac{\partial \phi(n, t)}{\partial t} = -\frac{\partial [M(n) \phi(n, t)]}{\partial n} + \frac{1}{2} \frac{\partial^2 [S(n) \phi(n, t)]}{\partial n^2} \quad (9a)$$

The corresponding Kolmogorov backward equation is then

$$\frac{\partial \phi(n, t)}{\partial t} = M(n) \frac{\partial \phi(n, t)}{\partial n} + \frac{S(n)}{2} \frac{\partial^2 \phi(n, t)}{\partial n^2} \quad (9b)$$

If another genotype with census number $n^*$ is introduced into the population, the corresponding equations are of course of the same form as (8) and (9), with n* in place of n. If one imposes a constraint of constant population size, such that n+n*=N (i.e. competition is such that there is a single carrying capacity for both genotypes), one can derive diffusion equations for the density function of the frequency p of the first allele, where p=n/(n+n*).

Diffusion equations for allele (or genotype) frequency are derived by applying the chain rule formalism for random variables proposed by Ito (see for example Karlin and Taylor 1975, pgs 347-8, Mikosch 1998), under the assumption of no implicit time dependence in f(n), and assuming zero covariance between n and n* (so that mixed second partials can be ignored),

$$\mathrm{df}(n, n^*) = \frac{\partial f}{\partial n} M(n) + \frac{\partial f}{\partial n^*} M(n^*) + \frac{1}{2}\left[\frac{\partial^2 f}{\partial n^2} S^2 + \frac{\partial^2 f}{\partial n^{*2}} S^{*2}\right] + \left(\frac{\partial f}{\partial n} \sqrt{S} + \frac{\partial f}{\partial n^*} \sqrt{S^*}\right) \mathrm{dB}_t$$

Applying this to $p(n,n^*) = n(n+n^*)^{-1}$, from which the first and second partial derivatives of f with respect to n and n* can be explicitly calculated, and setting N=n+n*, we have:

$$\mathrm{dp} = \frac{1-p}{N} r p N - \frac{p}{N}(1-p) r^* N +$$
$$\frac{1}{2}\left[\frac{(1-p)}{N^2} \sigma^{*2} N - \frac{p}{N^2} \sigma^2 N\right] +$$
$$\left(\frac{1-p}{N}\sqrt{\sigma^2 N p} - \frac{p}{N}\sqrt{\sigma^{*2} N (1-p)}\right) \mathrm{dB}_t \quad (10)$$

A formally identical derivation for diffusion approximations describing selection on variance in fitness (in the context of strategy-specific fedundities, without reference to age structure) was used by in Demetrius and Gundlach (2000).

The factor of the first partial term is the "drift" coefficient:

$$M(p) = \left(\Delta r - \frac{\Delta \sigma^2}{N}\right) p(1-p)$$

where $\Delta r = r - r^*$ and $\Delta \sigma^2 = \Delta \sigma^2 - \Delta \sigma^{*2}$.

The diffusion coefficient $S(p)$ can be calculated directly by taking second derivatives as defined by the Ito chain rule, specifically,

$$S(p) = \left(\frac{\partial f}{\partial n}\right)^2 S + \left(\frac{\partial f}{\partial n^*}\right)^2 S^*$$

which gives

$$S(p) = \frac{p(1-p)}{N}\left((1-p)\sigma^2 + p\sigma^{*2}\right)$$

The Kolmogorov backward equation for the probability density of allele frequency $\phi(p)$ with deterministic and Brownian motion coefficients corresponding to (10) is

$$\frac{\partial \phi(p)}{\partial t} = \left(\Delta r - \frac{\Delta \sigma^2}{N}\right) p(1-p) \frac{\partial \phi(p)}{\partial p} + \frac{p(1-p)}{2N}\left((1-p)\sigma^2 + p\sigma^{*2}\right) \frac{\partial^2 \phi(p)}{\partial p^2} \quad (11)$$

In the special case where the Malthusian parameters and the demographic variances of the two genotypes are equal, $M(p)=0$ and $S(p) = p(1-p)\sigma^2/N$, as derived in the appendix to Engen et al (2005b).

The above expressions for the drift and diffusion terms are of course formally equivalent to Gillespie's expressions (1a,b) if one substitutes demographic variance as a generalization of his "variance in offspring number". As in Gillespie's equations, the coefficient of the drift term is $(\triangle r - \frac{\triangle \sigma^2}{N})$ rather than $\Delta r$ (note that this argument can be readily generalized and extended to sexual diploids when fitness effects at different loci are additive or when the follow simple dominance relationships. This follows for the mean and variance in growth rate for the same reasons as for mean and variance in single age class fecundity analyzed by Gillespie).

It argued in Proulx (2000) and Shpak and Proulx (2007) that the sign of M(p) will determine whether an allele has a higher or lower than neutral probability of fixation, i.e. when E(Δp)>0, the probability of an allele with frequency p invading will be greater than its initial frequency, even in cases of strong selection where the diffusion approximations break down. If the contribution of demographic variance is significant, then using the expectation of the Malthusian parameter as a measure of fitness will be misleading, in that genotypes with lower values of r may actually be favored by selection as a consequence of selection on the second moments.

## Fixation and Invasion Probabilities: Analytical Predictions vs. Simulations

We first consider the case of two haploid, asexual genotypes with equal average growth rate, that differ in their demographic variances $\sigma_1^2$, $\sigma_2^2$. As was shown in Gillespie (1974), the diffusion estimate for fixation probability an allele of the first type (Equation 2) in this case reduces to

$$U(p) = \frac{\int_0^p ((1-x)\sigma_1^2 + x\sigma_2^2)^{-2}\,dx}{\int_0^1 ((1-x)\sigma_1^2 + x\sigma_2^2)^{-2}\,dx} = \frac{\sigma_2^2\, p}{(1-p)\sigma_2^2 + p\sigma_1^2} \qquad (12)$$

so that for $\sigma_1^2 > \sigma_2^2$, the probability of fixing the first genotype will be less than that of a neutral allele, i.e. less than its initial frequency p. This quantity is independent of population size, because while smaller N leads to a greater sample variance of growth rate, it also leads to decreased efficacy of directional selection.

For the life history matrix

$$A = \begin{pmatrix} 0 & 1 & 3 & 2 \\ 0.5 & 0 & 0 & 0 \\ 0 & 0.3 & 0 & 0 \\ 0 & 0 & 0.2 & 0 \end{pmatrix}$$

the expected growth rate $\lambda$=1.00389 (values close to unity were chosen so that population growth over multiple time steps and iterations would be minimal) and a demographic variance $\sigma_\lambda^2$=0.587288. On the log scale, we have r=0.00388 and $\sigma_r^2$=0.582747. The genotype with life history matrix **A** is put into competition with genotypes with the following life history matrices:

$$B = \begin{pmatrix} 0 & 2 & 6 & 4 \\ 0.25 & 0 & 0 & 0 \\ 0 & 0.3 & 0 & 0 \\ 0 & 0 & 0.2 & 0 \end{pmatrix},\ C = \begin{pmatrix} 0 & 3 & 9 & 6 \\ 0.125 & 0 & 0 & 0 \\ 0 & 0.3 & 0 & 0 \\ 0 & 0 & 0.2 & 0 \end{pmatrix},\ D = \begin{pmatrix} 0 & 4 & 12 & 8 \\ 0.125 & 0 & 0 & 0 \\ 0 & 0.3 & 0 & 0 \\ 0 & 0 & 0.2 & 0 \end{pmatrix}$$

which have the same expected growth rate as **A** but higher respective demographic variances (6), $\sigma_r^2$=1.13407, 1.73698, 2.15947.

These rather extreme differences in fitness variance arise as a consequence of the fact that in **D** versus **A** (for example), there is a fourfold decrease in survival probability from age class 1 to 2, compensated by a fourfold increase in age class fertilities. This leads to equal average growth rates but far greater variance in D, because a smaller fraction survive to reproductive age, while those that survive have many more offspring. It is unlikely that conspecific genotypes will have such drastic differences in their life history parameters, so the values are chosen to illustrate the strength of selection on demographic variance rather than to model an actual scenario of intraspecific competition.

Substituting these values into (2), one can calculate the expected fixation probability of a genotype with life history matrix **A** at initial frequency p in competition with **B**, **C**, or **D**. These predicted values are shown to correspond closely with the fixation probabilities estimated from individual based simulations. The simulations were implemented in *Mathematica* (the code is available from the author upon request). For a desired total population size N, two vectors with initial age distributions were specified such that the total number of individuals in the population summed to N. In competition between genotypes with an initial frequency of 0.5, both distributions were chosen to be as close to the stable age distribution (normalized first eigenvector) as possible. In invasion analysis, a "resident" population (background of genotypes with life histories given by **B**, **C**, **D**, etc.)

is chosen to be close to the stable age distribution while a single individual mutant with life history matrix **A** is represented by one individual in the first age class.

To iterate the process, each individual in a given age class either survives or fails to do so by the next age interval with probability $b_i$. In the simulations, this random process is represented by selecting a value from a binomial distribution Bin($x_i, b_i$), where $x_i$ is the number of individuals of a genotype in age class i. While it is assumed for simplicity that there is no inherent variance in offspring number (so that each individual of either genotype in the ith class produces exactly $m_i$ offspring), there is an additional parameter $b_0$ (set to unity in the above examples) which determines the probability of newborns surviving until the prereproductive time of census. This is simulated by choosing a random number of offspring at the census point with distribution Bin($m_i x_i$, $b_0$), which is done for both genotypes. After the reproduction and survivorship iterations are completed, the total population size is set to N by culling numbers of individuals from each genotype and age class in proportion to their representation in the next time step.

Every cycle of survival and birth is run for as long as is necessary for one of the genotypes to become lost or fixed. In turn, the simulations involve several thousand trial runs, each of which is iterated until fixation or loss, so that the average number of fixation events can be calculated from a sufficiently large number of trials. It is this value, and "empirical" estimate of fixation probability across simulations, that will be compared with the predicted value U(p).

For the simulations involving equal growth rates r, a large population size of N=1000 was chosen to minimize the sampling error effects due to culling. Since the fixation probability is independent of population size (even though the time until fixation scales linearly with N), there is no loss of generality in using a large value of N in these examples. Table 1 shows predicted versus observed fixation probabilities of the "**A**" genotype given initial frequencies p=0.5 and p=0.001 against **B**, **C**, and **D**. As can be seen, the agreement is quite close.

Constructing life history matrices such that the genotype with a higher growth rate also has a higher demographic variance is straightforward, and it's fairly obvious how a trade-off leading to this scenario might arise. For instance, an organism may sacrifice its survival probability across age classes in favor of producing a higher number of offspring in a given age class, leading to a significant difference in demographic variance with or without major changes in the growth rate.

For example, contrast matrix **A** with the following:

$$E = \begin{pmatrix} 0 & 2 & 6 & 5 \\ 0.25 & 0 & 0 & 0 \\ 0 & 0.3 & 0 & 0 \\ 0 & 0 & 0.2 & 0 \end{pmatrix},$$

$$F = \begin{pmatrix} 0 & 3 & 9 & 7 \\ 0.166667 & 0 & 0 & 0 \\ 0 & 0.3 & 0 & 0 \\ 0 & 0 & 0.2 & 0 \end{pmatrix}, \quad H = \begin{pmatrix} 0 & 4 & 12 & 9 \\ 0.125 & 0 & 0 & 0 \\ 0 & 0.3 & 0 & 0 \\ 0 & 0 & 0.2 & 0 \end{pmatrix},$$

which have respective Malthusian parameters r=0.009558, 0.0076845, 0.00674 (all slightly larger than $r_A$=0.003881), and demographic variances $\sigma_r^2$=1.1482, 1.6631,

2.1705 (significantly greater than 0.582747 for matrix A).

Equation (1a) predicts that for sufficiently small population sizes, genotype **A** will have a higher than neutral probability of fixation. Specifically, the critical population size at which the "effective fitness" of **A** is equal to that of **E**, and **H** is given by $N = \frac{\Delta \sigma^2}{\Delta r}$, which for the three competing genotypes is 100, 284, and 555. As a consequence, it is expected to see fixation probabilities approximately equal to initial frequency p when N equals these values. For N smaller than these critical values, it is predicted that strategy "**A**" will be favored over competitors due to its lower demographic variance, in spite of having a slightly lower growth rate, while the converse should be true for sufficiently large population sizes. The discrepancy between fixation probabilities at large versus small population sizes should of course be most pronounced when the difference in variance between strategies is greatest, is seen by comparing genotype with Leslie matrix **A** to the genotype with matrix **H**.

Figure 1a (solid line) plots the fixation probabilities of the "**A**" genotype given an initial frequency p=0.5 in competition against a genotype with life history **E**, for a range of population sizes N=50, 100, 200, 500, and 1000. Figure 1b does the same for a single invading mutant of type **A** against a background of **E**. To correct for the fact that the initial frequency, and consequently the neutral probability of fixation, is higher in small population (a single invading mutant has frequency 0.02 for N=50, 0.001 for N=1000), the fixation probabilities are multiplied by the total population sizes N. Therefore, while in Figure 1a values greater than 0.5 indicate greater than neutral fixation probabilities, in 1b values greater than unity indicate that low variance strategy **A** has a higher than neutral probability of invasion. Along the same lines, Figures 2a and 3a show fixation probability of **A** at 50% frequency against strategies **F** and **H**, while 2b and 3b show the corresponding invasion probabilities multiplied by population size N corresponding to initial frequencies of p=1/N.

Superimposed against the plots of fixation frequency are solutions to Equation (2) given the same growth and variance parameters as in each simulation (shown as dashed lines). The correspondence between predicted fixation probabilities U(p) and the frequencies in the simulations is frequently not as close as those in Table 1 (where the growth rates are equal), particularly in the smaller population sizes. Furthermore, in both Table 1 and in Figures 1-3, the difference between U(p) and the simulation frequencies is more pronounced in the invasion analysis, where p=1/N (a single invading genotype) than in competition experiments where p=0.5 (N/2 individuals of each genotype).

These discrepancies are a consequence of deviations from equilibrium age distributions. Working with the diffusion approximation (11) and its solutions involves reducing a description in terms of the number of individuals in each age classe to a description in terms of total census number irrespective of the age class distribution. Of course, this can only be done if an equilibrium age class distribution is closely approximated. For large population sizes and large initial numbers of both genotypes, the desired distribution can readily be approximated. However, when population sizes are small or the number of any genotype is low, one or both of the age class distributions (for the two genotypes) will be far from the equilibrium at which their growth rates and variances are estimated. This should be particularly obvious in the case of an invader, since by definition a single individual cannot be at or even closely approximate a stable age distribution! Only if the invader increases rapidly to a reasonable census number (or in special cases where most of the population consists of juveniles at equilibrium) should the univariate representation be expected to hold.

To understand how deviations from a stable age distribution can lead to misleading predictions of invasion probabilities, one must consider the population reproductive value of a genotype. For organisms with high juvenile mortality rates, the reproductive value of a newborn is almost always lower than that of the younger reproductive age classes. Therefore, the "population reproductive value" of a single juvenile will be lower than that of a population with a stable age distribution. This leads to a lower growth rate than predicted from the Leslie matrix (i.e. $\lambda$ overestimates the growth rate of the rare genotype), and consequently a lower probability of invasion than is given by U(p) based on r and $\sigma^2$ alone.

For example, consider an "**A**" mutant juvenile invading against a background of N-1 "**B**" individuals, where the latter are in an approximate equilibrium age distribution. The reproductive value of a juvenile "**A**" is slightly greater than unity, while the popula-tion reproductive value of **B** is 1.92. As a result, even though at equilibrium the expected growth rates are equal and the demographic variance of **A** is lower, there will be an initial bias against the invader until a critical density of later age classes is attained. This accounts in part for the invasion probabilities being in most cases lower than predicted from the diffusion approximation. Nevertheless, it should be emphasized that at least qualitatively the approximations seem to be quite robust. Even for small populations and a single invader, using growth rate and its variance in the diffusion model correctly predicts the population sizes at which the invader has a higher or lower than neutral fixation probability.

As an aside, it was noted above that the use of growth rate and demographic variance in describing the dynamics of N(t) is an approximation that depends on the use of variance in population number as a proxy for variance in reproductive value. To determine to what extent this approximation is valid, the selection dynamics were implemented over several (t=5, 10, etc) generations in the presence of a single genotype, such as the one given by matrix A. For multiple trials (1000 or more), the variance in N(t) was calculated. When the initial age distribution was close to $v$, the $\sigma^2$ was a good predictor for var[N(t)], within a margin of error of $10^{-3}$ when t was sufficiently large.

## Genetic Drift, Demographic Stochasticity, and Effective Population Size

In addition to demographic variance, there are other sources of stochasticity that drive changes in gene frequency, though their interaction with directional selection is quite different from intrinsic, intragenerational variance (Gillespie 1975, Lande et al 2003). Intuitively, the additivity of variances should allow one to combine different sources of variance in the coefficients of the diffusion and drift terms of the Kolmogorov equations.

Extrinsic variance due to environmental fluctuation was first formally investigated in Haldane and Jayakar (1963) who argued that when the fitness of a genotype fluctuates across generations, the geometric mean fitness is a better predictor of fixation probability than the arithmetic mean. When this variance term is small, it was shown in studies by Lewontin and Cohen (1969), Gillespie (1973), Turelli (1977), and Tuljapurkar and Orzack (1980) that the relevant selection coefficient (i.e. factor of E[Δp]) is approximately

$$\Delta r - \frac{\sigma_e^2}{2}$$

Environmental variance does contribute to the first moment of change in population number (and allele frequency). However, it is generally assumed that stochasticity in survival and reproduction due to environmental fluctuations are uncorrelated with genotype. As a result, the effective fitness loss due to environmental variance is the same for both genotypes, so that no fitness differential $\Delta \sigma_e^2$ is induced (in contrast to intrinsic demographic variance). Nevertheless, environmental variance (as does Fisher-Wright-Kimura genetic drift) introduces an additional source of stochasticity into the population dynamics that affects the probability of invasion and fixation. This can be seen by looking at the first and second moments of Δp.

For n individuals of a given asexual haploid genotype, the combined contribution of demographic and environmental variance to the first and second moments of dn is M(n)=rn and S(n)=$\sigma_d^2 n + \sigma_e^2 n^2$, with similar expressions for change in the number n* of a competing genotype (with corresponding growth and variance terms r*, $\sigma_d^{*2}$). It is assumed that environmental stochasticity influences the fitness of both genotypes in the same way, so that there is a single environmental variance term. Applying the Ito stochastic chain rule formalism, and the notation Δr=r-r*, $\Delta\sigma_d^2 = \sigma_d^2 - \sigma_d^{*2}$, and N=n+n*,

$$M(p) = \left(\Delta r - \frac{\Delta\sigma_d^2}{N} - \sigma_e^2 (1 - 2p)\right) p (1 - p)$$

and

$$S(p) = \frac{p(1-p)}{N} \left((1-p)\sigma_d^2 + p\sigma_d^{*2}\right) + p^2 (1-p)^2 \sigma_e^2$$

Again, there is no selection differential between genotypes due to environmental variance (which is why in models with extrinsic stochasticity, the difference between Malthusian parameters is sufficient to predict which genotype will have a higher than neutral probability of fixation, even in small populations). In this regard, the effects are similar to those of Wright-Kimura genetic drift due to gamete sampling, though in the latter the sampling variance only contributes to the second moment, i.e. under the assumption of sampling with replacement

$$E[(\Delta p)^2] = \frac{p(1-p)}{N}$$

so that the combined effects of the various sources of stochasticity (demographic, environmental, and gamete sampling) give us

$$M(p) = \left(\Delta r - \frac{\Delta \sigma_d^2}{N} - \sigma_e^2 (1 - 2p)\right) p (1-p)$$

$$S(p) = \frac{p(1-p)}{N} \left(1 + (1-p)\sigma_d^2 + p\sigma_d^{*2}\right) + p^2 (1-p)^2 \sigma_e^2$$

The first term in the expression for S(p) which combines classic genetic drift with demographic stochasticity, is equivalent to the derivation in Gillespie (1975), apart from the scaling terms and the interpretation of the demographic variance term. If environmental variance can be estimated and the life history matrix L is known, the above allows one to calculate fixation probabilities of genotypes in finite populations where all of the relevant deterministic and stochastic factors are taken into consideration. Namely, for the Kolmogorov backward equations (11), the solution for the estimate of fixation probability U(p) given initial frequency p is (e.g. Karlin and Taylor 1975, Crow and Kimura 1970, Ewens 2004)

$$U(p) = \frac{\int_0^p \mathrm{Exp}\left[-2 \int \frac{M(y)}{S(y)} dy\right] dx}{\int_0^1 \mathrm{Exp}\left[-2 \int \frac{M(y)}{S(y)} dy\right] dx}$$

In the cases where the only source of variance considered is demographic, the above has a closed form solution to (2), while introducing genetic drift gives

$$U(p) = \frac{\int_0^p (1 + (1-x)\sigma_1^2 + x\sigma_2^2)^{2\left(\frac{N(\Delta r)}{\Delta \sigma^2} - 1\right)} dx}{\int_0^1 (1 + (1-x)\sigma_1^2 + x\sigma_2^2)^{2\left(\frac{N(\Delta r)}{\Delta \sigma^2} - 1\right)} dx} \quad (13)$$

Combining this with environmental variance introduces leads to an expression for U(p) where the integrals cannot be explicitly evaluated (the "internal" integrals involve sums of arctangent and logarithmic functions). However, it is nevertheless possible to find numerical solutions and to approximate the fixation probabilities under different life history regimes and environmental fluctuations.

Focusing on the interaction of demographic variance and genetic drift alone, if one treats r-$\frac{\sigma^2}{N}$ as effective fitness, then genetic drift in this context plays the same role as it does in classical selection drift regimes. When $\Delta r - \frac{\Delta \sigma_d^2}{N} > 0$ the fixation probability of the reference genotype remains higher than neutral, but less so than would be predicted in the absence of finite gamete sampling (conversely when it is negative). For instance, consider competition between genotypes **A** and **H** again. For a population size N=100, in the absence of genetic drift proper, the fixation probability of genotype **A** with initial frequency p=0.5 is U(0.5)=0.765. If one includes drift, the fixation probability is U(0.5)=0.652. While still giving a higher than neutral fixation probability, the quantity is lower as a consequence of the additional source of stochasticity. The effect is expected to be most pronounced at small population sizes, in spite of the fact that lower values of N also lead to a strong demographic variance contribution to effective fitness.

The S(p) term also suggests an effective population size with respect to the contribution of the diffusion term. If Fisher-Wright-Kimura genetic drift were the only source of stochasticity in the change of allele frequency, the effective size (in the absence of spatial structure, inbreeding, etc) is equal to the census size N. Defining variance effective population number $N_e$ to be the population size that would induce the same stochasticity in the absence of demographic variance,

$$\frac{p(1-p)}{N}(1 + (1-p)\sigma_d^2 + p\sigma_d^{*2}) = \frac{p(1-p)}{N_e}$$

$$N_e = \frac{N}{(1 + (1-p)\sigma_d^2 + p\sigma_d^{*2})}$$

When the two genotypes have the same demographic variance, the above expression reduces to the effective population size calculated for haploids in Felsenstein (1971) and Hill (1972), if one makes the appropriate substitutions for variance in terms of generation time and reproductive value, i.e.

$$N_e = \frac{N}{(1 + \sigma_d^2)} =$$

$$\frac{N}{\left(1 + \sum_i \left[\left(\frac{\lambda^{-i+1}}{\tau}\right)^2 \text{var}[F_i] + \left(\frac{\lambda^{-i} l_i V_{i+1}}{\tau}\right)^2 \frac{b_i(1-b_i)}{V_i}\right]\right)}$$

## Discussion

Calculating the mean and variance of growth rate (or of the Malthusian parameter on a logarithmic scale) allows one to describe what was a multidimensional problem (with the number of variables corresponding to d life history stages) into a univariate dynamical system n(t). Furthermore, in the case of two genotypes, writing down coupled equations for n(t) and n*(t) and applying the appropriate transforms allows one to approximate the stochastic dynamics of p(t), the frequency of individuals of the first type in the population, in terms of the mean and variance differences in growth rate. This bivariate problem can be reduced to a univariate one as well if one posits a model of competition where genotype carrying capacities are equal and total population size is fixed.

Since demographic variance can be considered an extension of Gillespie's fecundity variance to age structured populations, it is to be expected that in a finite population, the mean growth rate $\lambda$ will not generally be a sufficient predictor of whether a genotype with a given life history can invade and tend to fixation. This is due to the fact that demographic stochasticity (and growth rate variance generally) relates to population fluctuations and increased extinction probabilities (Lande et al 2003, Engen et al 2005a), a result that applies to multiple competing lineages as well as to the genetically monomorphic (with respect to growth rate and variance) populations studied by Lande, Engen and colleagues. Moreover, unlike extrinsic environmental stochasticity, demographic stochastic-ity depends on the specific parameters of a genotype's life history, and therefore contributes to the selection coefficient of the expected allele frequency change when two genotypes are in competition. This suggests an effective selection differential measure

$$s \approx \Delta r - \frac{\Delta \sigma^2}{N}$$

and an "effective fitness" coefficient $r - \frac{\sigma^2}{N}$ for a single genotype. Since in a finite population the fixation probabilities of competing genotype depend both on the growth rate and demographic variance (insofar as a genotype with higher effective fitness have a higher than neutral probability of invasion), it is necessary to consider the effects on both first and second moments in evaluating sensitivity of "fitness" to changes in individual life history parameters.

In strictly deterministic analyses of selection for life history parameters, the sensitivity of growth rate $\lambda$ with respect to age-specific parameters $b_i$ and $F_i$ was evaluated as the partial derivative of r or $\lambda$ with respect to $b_i$ or $F_i$ (e.g. Hamilton 1966, Demetrius 1969). In finite populations, it may be more informative to evaluate partials of effective fitness, i.e.

$$\frac{\partial \left(r - \frac{\sigma^2}{N}\right)}{\partial b_i}, \quad \frac{\partial \left(r - \frac{\sigma^2}{N}\right)}{\partial F_i}$$

For details on calculating these partial derivatives (which are rather tedious for the variance terms), the reader is referred to the Appendix.

The importance of such measures hinges on the extent of the contribution of demo-

*DemStochGenetics.nb* 34graphic variance in determining fixation probabilities. If the demographic variance itself is negligible, as would be the case in a life history where most individuals survive up to an age where most of the reproduction occurs (with little variance in offspring output per age class or individual), then a strictly deterministic approximation based on measures of mean growth rates should correctly predict the outcome of selection dynamics. When the demographic variance is not negligible (meaning that the difference in growth rate variance between two genotypes is greater than the difference in mean growth rate by a magnitude that exceeds N), the importance of the variance contribution will depend on the population size. If the population size is very large (as shown in Figures 1-3, for N=1000), the fixation or invasion probabilities are determined almost entirely by the difference in growth rate, while for small population sizes (e.g. N=50), the variance is the principal determinant of invasion probability.

    Traditionally, the grounds for neglecting variance in fitness (with or without age structure) has been the fact that the census population number of most model organisms tends to be large, often numbering in the thousands for larger organisms such as birds and mammals and in the millions for insects and small marine invertebrates. For such population sizes, values of $\Delta\sigma^2/N$ of comparable order to $\Delta r$ are extremely unlikely, since the difference in demographic variance between genotypes of the same species tends not to be very large. In the simulation examples analyzed, extreme (and biologically unrealistic) differences in variance were induced by having 2 to 4-fold differences in survival rate

compensated with equalfold differences in age class fecundities. Such matrices in were chosen to illustrate the efficacy of selection against demographic variance rather than reflect biological realism.

If one assumes that the majority of mutations lead to small changes in life histories (i.e. somewhat greater degrees of iteroparity and semelparity rather than extreme changes), one should generally expect $\Delta\sigma^2$ to be of not much greater order of magnitude than $\Delta\lambda$. In such a case, except when population sizes are extremely small, selection on variance is not expected to be sufficiently strong to favor the genotype with lower growth rate when there are trade-offs between mean and variance in fitness, nor even to noticeably change fixation probabilities of the favored genotype from the values of U(p) predicted by standard selection and drift equations.

However, recent work (Shpak and Proulx 2007, Lehmann and Balloux 2007) suggests that for subdivided populations, census size is not an adequate predictor for the strength of selection on variance in offspring number, at least not under a regime of soft selection (sensu Wade 1984, where most of the selective culling of individuals takes place in small individual demes prior to migration, as would be the case in large social vertebrate animals and other taxa where offspring remain in local family groups until the age of sexual maturity). If the migration across demes in a metapopulation occurs among sexually mature adults, then the census size $n_d$ of each deme (as opposed to the census size N of the metapopulation) is a better predictor for the sample variance in offspring number. This means that even for very large total N, if the individual demes are small (for instance, $n_d$ of order 10) selection against variance in offspring number can be significant. Indeed, it was shown that this result is entirely independent of migration rate in the extreme case of "pure" soft selection.

In the case of hard selection (where migration occurs among juveniles prior to culling), it was shown that the effective population size, and therefore the outcome of selection on variance, will depend strongly on the extent of migration. Low migration rates give sample variance that scale approximately as n, approaching N for higher rates. Specifically, in the case of approximately equal allele frequencies in each deme, the deme size relevant to fecundity sample variance is

$$n_e \approx \frac{n_d D}{D(1-m)^2 + m^2},$$

which approximates $n_d$ for low migration rates and nD=N with complete mixing.

In other words, even in seemingly large populations the contribution of intrinsic growth rate, variance can be significant enough to alter the outcome of selection, when migration rates are low or when selection is . Since this paper shows that the demographic variance contributions to the dynamics of N(t) and p(t) in age structured populations in the same way as offspring number (fecundity) variance in populations without age structure, the results for metapopulations should be robust for systems where both age and spatial structure are considered.

Indeed, an explicit model of age structure would allow one to investigate different degrees of hard and soft selection through the use of matrix models that combine age classes with migration between demes. By having different migration rates specific to different age classes, hard and soft selection become part of the spectrum rather than absolutes. For instance, a regime where pre-reproductive juveniles migrate at a relatively higher rate will tend to be more "hard" while the converse scenario more "soft." It would be desirable to investigate how changes to life history and migration parameters affect the outcome of selection in such a system.

## Acknowledgements

The author wishes to thank Russell Lande and Steinar Engen for clarifying the results and derivations in their original papers on demographic variance. I also thank Stephen Proulx, Lloyd Demetrius, and Michael Bulmer for helpful discussions of various aspects of life history evolution, as well as Ernie Barany and Robert Smits for advice on stochastic calculus and diffusion approximations. This research has been supported by startup funds for new faculty at the University of Texas at El Paso.

## Tables and Figures

# Table 1:

*Expected versus observed fixation probabilities of "**A**" versus other life histories*

|  | $p = 0.5$ | $p = 0.001$ |
|---|---|---|
| $\Delta\sigma_B^2$ | 0.661 | $1.94 \times 10^{-3}$ |
|  | 0.641 | $1.24 \times 10^{-3}$ |
| $\Delta\sigma_C^2$ | 0.739 | $2.82 \times 10^{-3}$ |
|  | 0.735 | $2.40 \times 10^{-3}$ |
| $\Delta\sigma_D^2$ | 0.787 | $3.70 \times 10^{-3}$ |
|  | 0.790 | $3.96 \times 10^{-3}$ |

For competition of the "**A**" genotype against **B, C**, and **D**, the respective fixation probabilities (upper value) calculated from the diffusion equation are shown in comparison with the fixation frequencies in the simulations (lower values). Fixation probabilities in the case of effective neutrality equal the initial frequency p.

# Figure 1a:

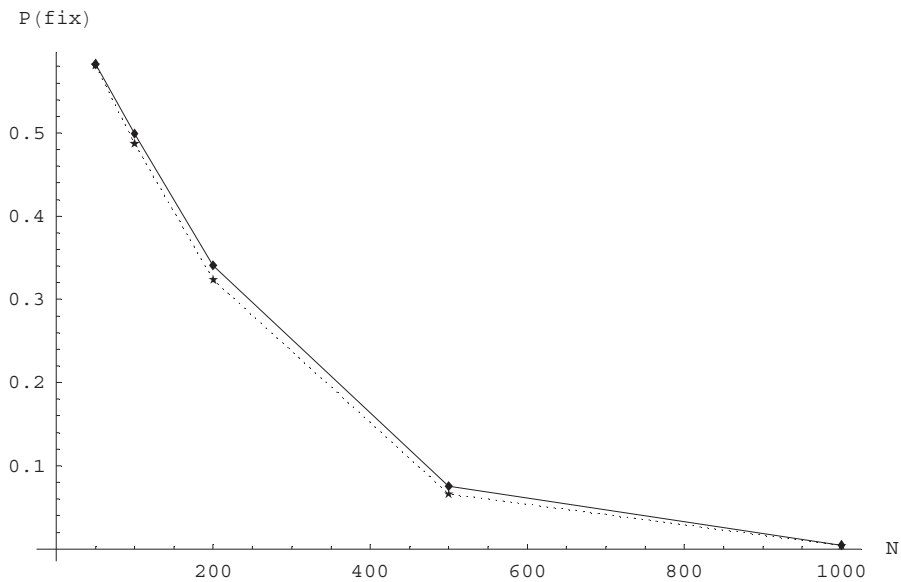

Fixation probabilities of the "**A**" asexual, haploid genotype against "**E**" genotype is plotted over a range of population sizes N, given initial frequency p=0.5. The solid line (with points at N=50, 100, 200, 500, and 1000) shows fixation frequencies from simulations, the dashed line the value of U(p) calculated for the same parameters.

*DemStochGenetics.nb* 43# Figure 1b:

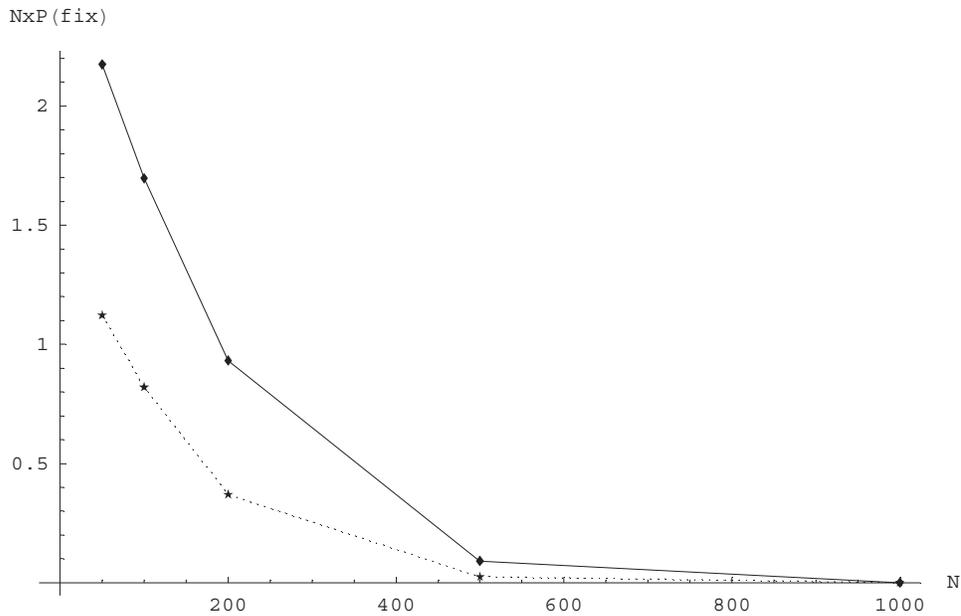

Invasion probabilities (multiplied by population size N for scale) of the "**A**" genotype against a background of **E** genotypes is plotted over a range of population sizes N, given initial frequency p=1/N. As in 1a, the solid line has points from simulations, the dashed line is based on solutions to the Kolmogorov backward equations.

# Figure 2a:

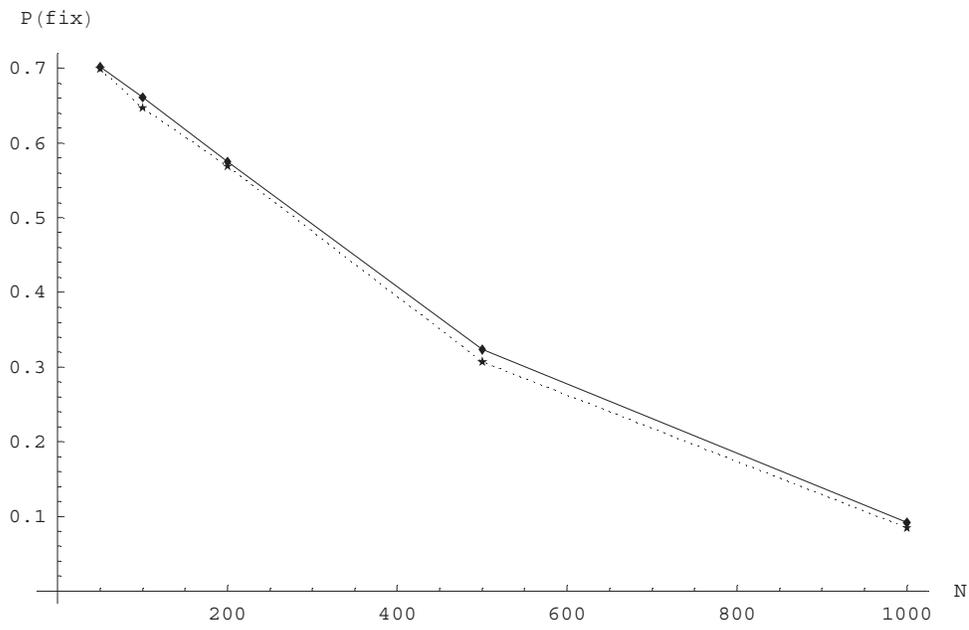

Same as 1a, but with competition between **A** and **F** genotypes.

# Figure 2b:

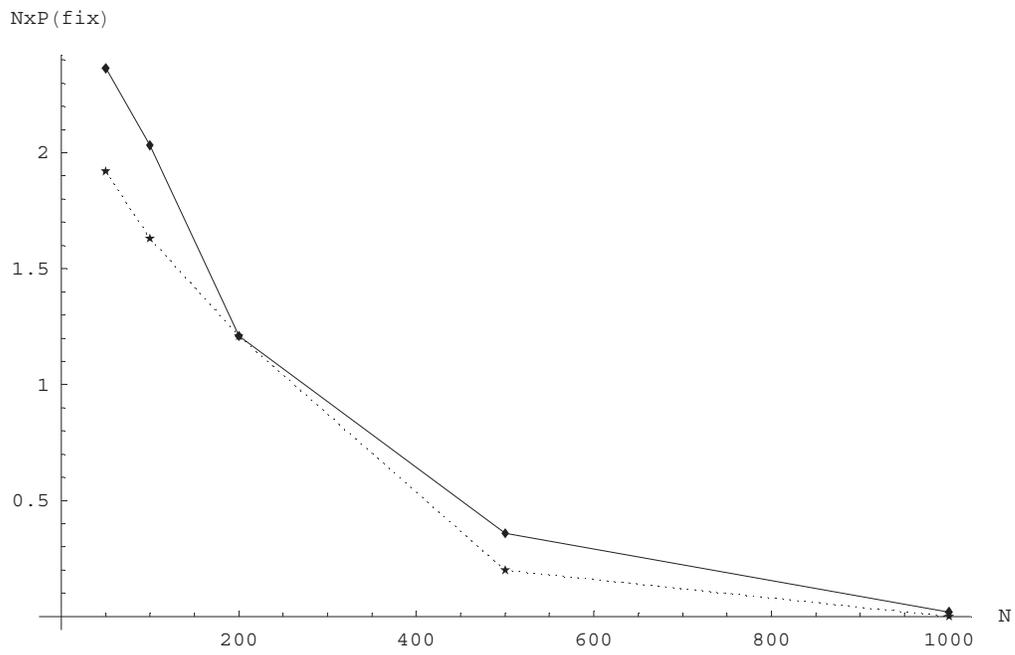

Same as 1b, but with invasion of **A** against a background of N-1 **F** genotypes.

# Figure 3a:

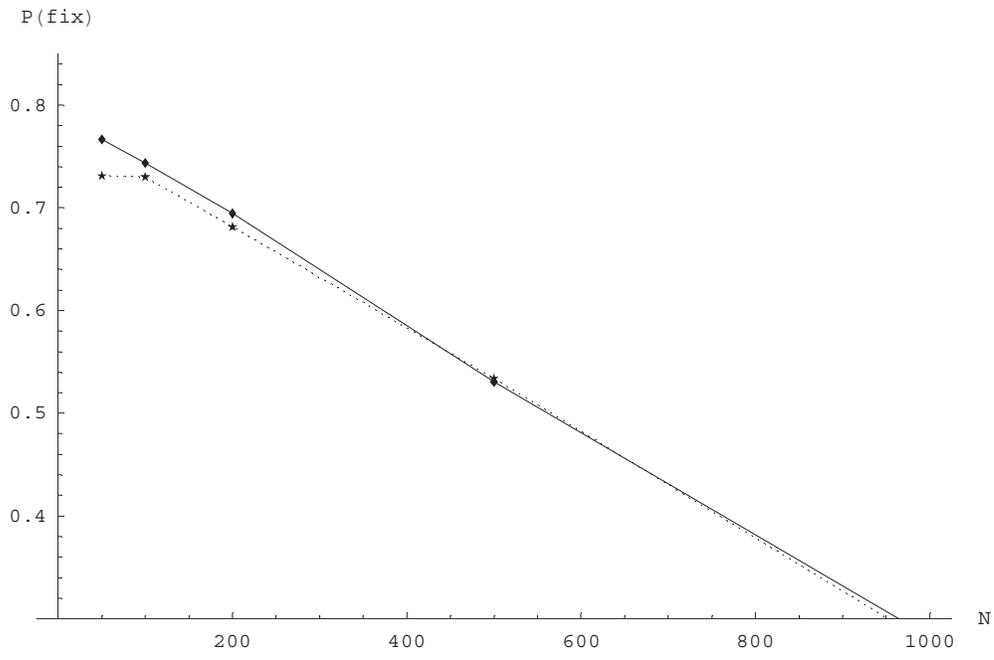

Same as 1a, but with competition between **A** and **H** genotypes.

# Figure 3b:

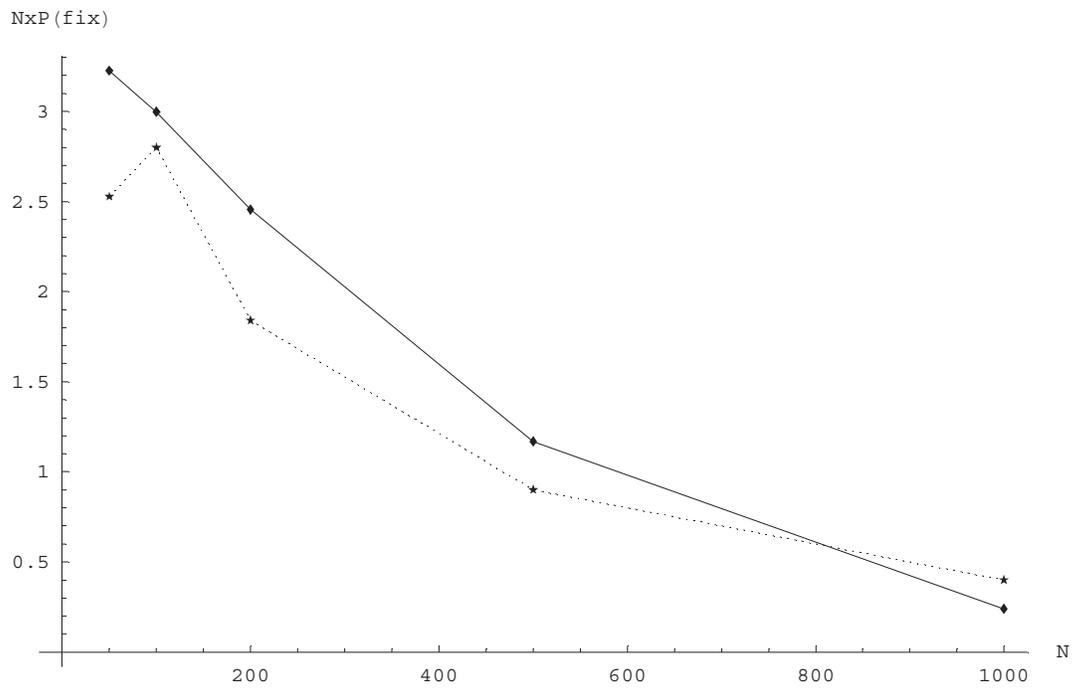

Same as 1b, but with invasion of **A** against a background of N-1 **H** genotypes.

## Appendix: Sensitivity of Selection Coefficient

It is suggested that for finite populations, evaluating the partial derivatives of r-$\frac{\sigma}{N}$ with respect to survival and fertility values provides a more useful measure of the sensitivity of "fitness" to changes in individual life history parameters. By evaluation partials of s with respect to b and f, we will measure the elasticity of effective fitness and estimate the strength of selection on individual b and F values when population size N is small.

$$\frac{\partial (r - \frac{\sigma^2}{N})}{\partial b_i} = \frac{1}{\lambda} \frac{\partial \lambda}{\partial b_i} - \frac{1}{\lambda^2} \frac{\partial \sigma^2}{\partial b_i}$$

$$\frac{\partial (r - \frac{\sigma^2}{N})}{\partial F_i} = \frac{1}{\lambda} \frac{\partial \lambda}{\partial F_i} - \frac{1}{\lambda^2} \frac{\partial \sigma^2}{\partial F_i}$$

Of course, the derivatives of growth rate $\lambda$ are known,

$$\frac{\partial \lambda}{\partial b_i} = \frac{\lambda^{-i} l_i V_{i+1}}{\tau}, \quad \frac{\partial \lambda}{\partial F_i} = \frac{\lambda^{-(i-1)} l_i}{\tau}$$

The contribution from the variance terms are

$$\frac{\partial \sigma_\lambda^2}{\partial b_j} =$$

$$\frac{\partial}{\partial b_j} \left[ \sum_i \left[ \left( \frac{\partial \lambda}{\partial F_i} \right)^2 \mathrm{var}[F_i] + \left( \frac{\partial \lambda}{\partial b_i} \right)^2 \frac{b_i (1 - b_i)}{v_i} \right] \right]$$

=

$$\frac{\partial}{\partial b_j} \left[ \sum_i \left[ \left( \frac{\lambda^{-i+1}}{\tau} \right)^2 \mathrm{var}[F_i] + \left( \frac{\lambda^{-i} l_i V_{i+1}}{\tau} \right)^2 \frac{b_i (1 - b_i)}{v_i} \right] \right]$$

Evaluating the partial derivates, we obtain

$$\sum_i \left[ 2 \operatorname{var}[f_i] \left( \frac{\partial \lambda}{\partial F_i} \right) \left( \frac{\partial^2 \lambda}{\partial b_j \partial F_i} \right) + \right.$$

$$\left. 2 \left( \frac{\partial \lambda}{\partial b_i} \right) \left( \frac{\partial^2 \lambda}{\partial b_j \partial b_i} \right) \frac{b_i (1 - b_i)}{v_i} \right] -$$

$$\sum_i \left( \frac{\partial \lambda}{\partial b_i} \right)^2 \frac{b_i (1 - b_i)}{v_i^2} \frac{\partial v_i}{\partial b_i} + \left( \frac{\partial \lambda}{\partial b_i} \right)^2 \frac{(1 - 2 b_j)}{v_i}$$

where the stationary distribution $v_i$ is

$$v_i = \frac{l_i \lambda^{-i}}{\sum_k l_k \lambda^{-k}}$$

so we substitute the following expression for its derivative in the last term of (??)

$$\frac{\partial v_i}{\partial b_j} = \frac{l_i \lambda^{-i}}{b_j \sum_k l_k \lambda^{-k}} - \frac{i \, l_i \lambda^{-i-1}}{\sum_k l_k \lambda^{-k}} \left( \frac{\partial \lambda}{\partial b_j} \right) -$$

$$\frac{l_i \lambda^{-i}}{(\sum_k l_k \lambda^{-k})^2} \left( \frac{l_i \lambda^{-i}}{b_j} - i \, l_i \lambda^{-i-1} \left( \frac{\partial \lambda}{\partial b_j} \right) \right)$$

if j≥i and

$$\frac{\partial v_i}{\partial b_j} =$$

$$\frac{l_i \lambda^{-i}}{(\sum_k l_k \lambda^{-k})^2} \left( i \, l_i \lambda^{-i-1} \left( \frac{\partial \lambda}{\partial b_j} \right) \right) - \frac{i \, l_i \lambda^{-i-1}}{\sum_k l_k \lambda^{-k}} \left( \frac{\partial \lambda}{\partial b_j} \right)$$

otherwise. The partials of λ with respect to b and f are defined above, so second partials can readily be calculated and substituted.

In the case of partials with respect to fertility parameters f, we have

$$\frac{\partial \sigma_\lambda^2}{\partial f_j} =$$

$$\frac{\partial}{\partial f_j} \left[ \sum_i \left[ \left( \frac{\partial \lambda}{\partial f_i} \right)^2 \operatorname{var}[F_i] + \left( \frac{\partial \lambda}{\partial b_i} \right)^2 \frac{b_i (1 - b_i)}{v_i} \right] \right] =$$

$$\sum_i \left[ 2 \, \text{var}[f_i] \left( \frac{\partial \lambda}{\partial F_i} \right) \left( \frac{\partial^2 \lambda}{\partial F_i^2} \right) + \right.$$

$$\left. 2 \left( \frac{\partial \lambda}{\partial b_i} \right) \left( \frac{\partial^2 \lambda}{\partial b_j \, \partial F_i} \right) \frac{b_i \, (1 - b_i)}{v_i} \right] -$$

$$\sum_i \left( \frac{\partial \lambda}{\partial b_i} \right)^2 b_i \, (1 - b_i) \, \frac{\partial v_i}{\partial F_i}$$